\documentclass{emulateapj}
\def\etal{{et~al. }}

\shorttitle{Forming Nuclear Star Cluster}

\shortauthors{D. Andersen et al.}

\begin{document} 

\title{Caught in formation: the nuclear-cluster-to-be in NGC~2139 \footnote{Based on observations collected at the European Southern Observatory, Chile, proposal No. 74.B-0264}}

\author{David R. Andersen\altaffilmark{1}, C. Jakob Walcher\altaffilmark{2,3}, 
Torsten B\"oker\altaffilmark{4}, Luis C. Ho\altaffilmark{5}, 
Roeland P. van der Marel\altaffilmark{6}, Hans-Walter Rix\altaffilmark{7}, 
Joseph C. Shields\altaffilmark{8}}

\affil{
\altaffilmark{1}NRC Herzberg Institute of Astrophysics, 5071 West Saanich Road,
Victoria, BC, V9E~2E7, Canada, \\
\altaffilmark{2}Observatoire Astronomique de Marseille Provence, Traverse du Siphon, 13376 Marseille Cedex 12, France, \\
\altaffilmark{3}Institut d'Astrophysique de Paris, 98bis, Boulevard Arago, 75014 Paris, France, \\
\altaffilmark{4}ESA/ESTEC, Keplerlaan 1, 2200 AG Noordwijk, Netherlands, \\
\altaffilmark{5}The Observatories of the Carnegie Institution of Washington, 813 Santa Barbara Street, Pasadena, CA 91101-1292, USA, \\
\altaffilmark{6}Space Telescope Science Institute, 3700 San Martin Drive, Baltimore, MD 21218, USA, \\
\altaffilmark{7}Max Planck Institut f\"ur Astronomie, K\"onigstuhl 17, 69117 Heidelberg, Germany, \\
\altaffilmark{8}Department of Physics and Astronomy, Clippinger Research Laboratories 251, Ohio University, Athens, OH45701-2979, USA \\
}
\email{david.andersen@nrc-cnrc.gc.ca}

\begin{abstract}

Close to its center, the bulgeless galaxy NGC~2139 hosts 
a star cluster that is younger 
and less massive than any actual nuclear star cluster (NC) studied so far. 
We have measured the H$\alpha$ velocity field around the photometric center
of this galaxy using the VLT ARGUS integral field unit and GIRAFFE spectrograph
in order to constrain different proposed theories of NC
formation. We observe that the best-fit kinematic center and the 
candidate NC appear to be separated by 2.8\arcsec\ (320 pc). Indeed,
the kinematic center also is offset from the galaxy's photometric center
and a possible bar or extended region of star formation in which
the young cluster resides,
implying that this galaxy is not in dynamic equilibrium.
The H$\alpha$ flux map also reveals
other regions of strong star formation in the possible bar.
These observations suggest that a nascent NC is forming away from the
kinematic center  of NGC 2139 which may come to rest there on
a time scale of a few 100 Myr.

\end{abstract}


\section{Introduction}
\label{s:intro} 

Galaxy centers continue to attract special interest 
as they host a number of distinctive 
phenomena, such as active galactic nuclei, central starbursts 
and extremely high stellar densities. The last decade has shown 
that the evolution of galaxies is closely linked to the evolution 
of their nuclei, as evidenced by a number of relations between
global and nuclear properties 
(e.g. Magorrian \etal 1998; Ferrarese \& Merrit 2000; Gebhardt \etal 2000; 
Graham \etal 2001; Ferrarese 2002; H\"aring \& Rix 2004). 

In view of this general paradigm and as a contribution to the full census of 
galaxy nuclei over all Hubble types, we are studying the central region of 
late-type, bulgeless spirals. Prime candidates for the 
nuclei of bulgeless spirals have been identified 
in the form of ``nuclear star clusters'' (NCs). 
Such compact, photometrically distinct NCs are found in, or very near, the 
centers of spirals across all Hubble types (Phillips \etal 1996; 
Carollo, Stiavelli \& Mack 1998; B\"oker \etal 1999; Matthews \etal 1999; 
B\"oker \etal 2002, Balcells \etal 2003, Scarlata \etal 2004), as well 
as ellipticals (see e.g. C\^ot\'e \etal 2006 and references therein). 

Due to the lower surface brightness of the background galaxy, NCs in bulgeless 
galaxies can be studied in more detail than those in earlier Hubble types. 
It has been shown that NCs in bulgeless galaxies have a number of unusual 
properties: they often are the most luminous cluster of their host galaxy and 
they lie inside the errorbars assigned to the location of the photometric center 
(B\"oker \etal 2002). Although they are as compact as globular clusters 
(B\"oker \etal 2004), they are one order of magnitude 
more massive than the massive end of the Galactic globular cluster 
mass function (Walcher \etal 2005). They also show evident signs 
of repetitive star formation (Walcher \etal 2006), which may lead to 
the formation of disks that are photometrically distinct (bluer) from 
the underlying NC (Seth \etal 2006). Finally, there are two 
examples of AGNs hosting NCs: NGC~4395 (Fillippenko \& Ho, 2003) 
and NGC~1042 (Shields \etal 2008). There also is evidence that the mass 
of the NC correlates with the mass of its host galaxy (Rossa et al. 2006). 
NCs are thus prime candidates for 
representing the unique centers of otherwise bulgeless galaxies.

At least three different scenarios for the formation (and evolution) 
of NCs can account for most of their properties.

(1) NC formation may be a generic property of late-type 
spirals, if the dynamical center of the galaxy is an a priori
well-defined location. This would e.g. be the natural consequence of a dark 
matter halo with a cuspy density profile. Magneto-rotational 
instability could then produce a steady gas inflow onto the galaxy center 
(Milosavljevi\'c 2004). Recently, compression of gas through 
tidal forces has also been suggested as a possible formation route 
for NCs (Emsellem \& van der Ven, 2008). 
An observational signature of this scenario would 
be the coincidence of the NC location with the location of the 
kinematic center as well as a rather ordered velocity field.

(2) On the other hand NC formation could be a random process. 
In this picture, any randomly formed, ``free-roaming'' seed cluster close 
to the overall center of a bulgeless galaxy would 
accumulate further gas in the gas-rich central region. The 
deeper potential well of the cluster then induces star formation in the 
accreted gas, thus leading to the very compact, massive objects we observe.
The velocity fields would not necessarily be centered on the NC 
or be well-ordered. This scenario would be expected, if the central potential 
of bulgeless galaxies were similar to the constant density cores advocated 
by some in the literature (see e.g. de Blok, Bosma \& McGaugh, 2003 and 
references therein).

(3) Finally, objects broadly similar to NCs could  coalesce through the merging
of several young clusters formed close to each other in a starburst event
(Oh \&  Lin 2000; Fellhauer \etal 2002; Capuzzo-Delcetta \& Miocchi 2008).

However, by construction, the distinction between the three formation
mechanisms will be blurred with time. In either the second or third
formation mechanisms, 
the formed cluster is drawn  to the center of the potential well by
dynamical friction.
For the case of a deep central potential, 
the dynamical friction timescale (Chandrasekhar 1943) 
for a proto-NC can be estimated
from Equation (3) of Milosavljevi\'c (2004), which
gives a lower bound for the decay timescale $t_{\mbox{decay}}$
with which a cluster would sink to the kinematic center of the galaxy:
\begin{equation}
t_{\mbox{decay}} \gtrapprox 3 \times  10^7 \mbox{yr} \left( \frac{v_{\mbox{circ}}}
{50~\mbox{km s}^{-1}} \right) \left( \frac{r}{100 \mbox{pc}} \right)^2  \left(
\frac{M_{\mbox{cl}}}{10^6 \mbox{M}_{\odot}} \right)^{-1},
\label{eq:milos}
\end{equation}
where $v_{\mbox{circ}}$ is the velocity of the cluster on an assumed
circular orbit, $r$ is the radius of that orbit and $M_{\mbox{cl}}$ is
the mass of the cluster.
This decay timescale can be quite short, after which all three scenarios
follow the same evolutionary path, where a massive pre-existing cluster
soaks up any infalling gas.

Distinguishing between the three scenarios quoted above is not only central 
to the identification and formation of NCs, it also helps inform the debate
over the form of the central potential 
of low surface brightness galaxies 
(e.g. de Blok 2005 and Valenzuela \etal 2007). If scenario 
1) could be shown to be valid, this would strongly support the theory
predicting diverging mass profiles in bulgeless disc galaxies.
Scenarios 2) or 3) would be unlikely 
if the environment close to the NC were influenced by a steep gravitational
potential.

Observational signatures of any of the three NC formation mechanisms 
described above will most easily be found in galaxies with a young NC.
In particular, the velocity field around a NC is interesting, as it 
allows us to search for a possible offset between the kinematic center 
and the NC. Such an offset would allow to clearly distinguish between the
different formation scenarios. To maximize the chances of finding a
NC offset from the kinematic center, one would ideally target a galaxy
whose photometric center were offset from its NC.
In NGC 2139, we have a galaxy that meets these criteria: 
NGC 2139 contains a candidate NC that is somewhat
offset from the photometric center and 
is very young as compared to other NCs. This candidate NC is also less massive 
than any other NC whose mass has been measured.

After introducing the properties of NGC 2139 and its central cluster,
we present the H$\alpha$
emission line velocity field of NGC 2139
measured using the ARGUS integral field unit (IFU) of ESO's Very Large
Telescope (VLT),
focusing on the innermost region directly around the 
potential nucleus of the galaxy.  Finally, we evaluate our results
both in terms of the NC formation scenarios described above and in
terms of the possibility that the central cluster of NGC 2139 is 
actually a very young super star cluster (SSC) similar to those 
found in some starburst galaxies which may over time fade
from view and not become a galaxy proto-nucleus.

\section{Previous observations of NGC 2139}
\label{s:prev}

NGC 2139 is morphologically classified as an SABcd galaxy (de Vaucouleurs 
\etal 1991, RC3). It is one of the 1000 brightest galaxies in the HIPASS 
survey (Koribalski \etal 2004), and 
has a recession velocity of 1837 km/s and a HI velocity width of 
$W_{50}=206.7$ km/s. Based on its velocity width, it is expected to have
a maximum, projected rotation velocity of 94 km/s (Paturel \etal 2003).  
However, the HI profile available from the HIPASS archive is quite 
asymmetric with the primary peak having a velocity of $\sim1785$ km/s.  
A survey of spatially resolved 1.425 GHz emission from IRAS 
bright galaxies (Condon \etal 1996) indicates that the radio emission from
NGC 2139 comes from two components: a primary lobe associated with 
the main body of the galaxy and a secondary lobe associated with the plume 
extending south of the galaxy.
The large-scale morphology as seen from publicly available Digitized Sky
Survey (DSS)\footnote{The Digitized Sky Surveys were produced at the Space
Telescope Science Institute under U.S. Government grant NAG W-2166. The 
images of these surveys are based on photographic data obtained using 
the Oschin Schmidt Telescope on Palomar Mountain and the UK Schmidt 
Telescope. The plates were processed into the present compressed digital 
form with the permission of these institutions.} images 
is irregular showing a few pronounced spiral arms and some tidal features.
This, along with many HII-region candidates concentrated near the center, 
are evidence for strong star formation and a possible recent merger event. 
The HST WFPC2 F814W image (B\"oker \etal 2002)  
clearly shows a luminous band (again presumably star-forming) 
going through the center of the galaxy, indicating a possible bar or star
formation filament 
triggered through gas inflow to the center of the 
galaxy. This band also includes the luminous star cluster that
B\"oker \etal (2002) classified 
as a NC.

This central star cluster is unique in several ways. 
It is the youngest ($4.1 \times 10^7$ yr) star cluster in a sample of nine 
NCs studied in Walcher \etal (2006, see also Rossa \etal 2006). 
From an in-depth analysis of its spectrum these authors find 
that the spectrum is actually consistent with the cluster not containing 
any old stellar population whatsoever.
It also is the least massive in their
sample (Walcher \etal 2005), with a dynamically determined mass of just
$M_{\mbox{cl}} = 8.3 \times 10^5 \mbox{M}_{\odot}$.
We note that this star cluster is probably less massive than most NCs, as
Rossa \etal (2006) find from a much larger sample with masses derived from 
stellar population analysis that the inability to infer spectroscopically 
the populations of faint clusters does introduce a bias 
towards younger ages, but not necessarily towards higher masses. 
It is also noteworthy in this context that the effective radius
of the NGC~2139 central cluster is 10pc, making it one of the 
largest in the Boeker \etal (2004) sample.

\begin{figure*}
\vbox to 2.5in{\rule{0pt}{6.5in}}
\includegraphics{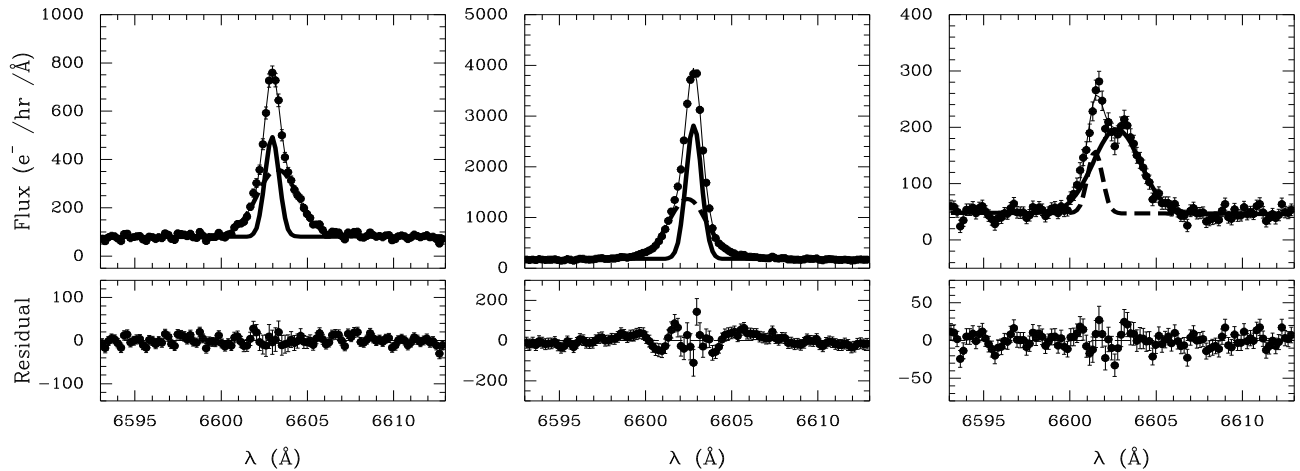}
\caption{The H$\alpha$ emission lines from three ARGUS IFU spaxels
with the best double-Gaussian fits (top) and fit residuals (bottom).
H$\alpha$ was detected in 298/299 spaxels, but
most line profiles cannot be well-fit by a single Gaussian.  Rather, almost
all (278/298) line profiles are better fit with a double Gaussian profile.
In the two panels towards the left, one component appears to be
stronger and have a narrower width while the other component fits the broad
wings.  The leftmost panel is an example of a type (2) profile identified in
the text.  The central panel shows a type (3) profile in which the broad wings
are offset from the
narrower peak. On the right is an example of a line with two clear kinematic
components (type 4, as identified in the text).  This latter phenomenon
was detected in 88/298 ARGUS spectra.
}\label{f:doubleline}
\end{figure*}

Indeed, as Walcher \etal (2005) posit, the central cluster in NGC 2139 
may be a forming NC,
thereby providing an excellent chance to identify an example for scenario 2).
Because the fate of the NGC 2139 cluster is still unknown, 
we will use the term ``central cluster'' (\emph{not} NC) for the remainder
of this paper, in order to avoid confusion with bona-fide NCs that (already)
occupy the kinematic center of their host galaxy.

\section{Data and Reductions}
\label{s:data}

Observations of NGC 2139 were carried out with the 
ARGUS integral field unit (IFU) coupled to the GIRAFFE spectrograph 
on the Kueyen telescope of ESO's Very Large Telescope (VLT) in May of 2004.
We used the 0.52\arcsec\ lenselet scale-size, i.e.
a field of view of 11.5\arcsec\ by 7.3\arcsec, and 
the LR06 grating which yields a resolution of $R\approx$13700 covering 
a wavelength range from 6400\AA $< \lambda\lambda <$ 7100\AA. 
The seeing during the observations was reported to be 0.8\arcsec.  Because
the goal of the observations was simply to study the velocity field, no
flux calibration was performed.

\begin{figure*}
\vbox to 2.8in{\rule{0pt}{6.5in}}
\includegraphics{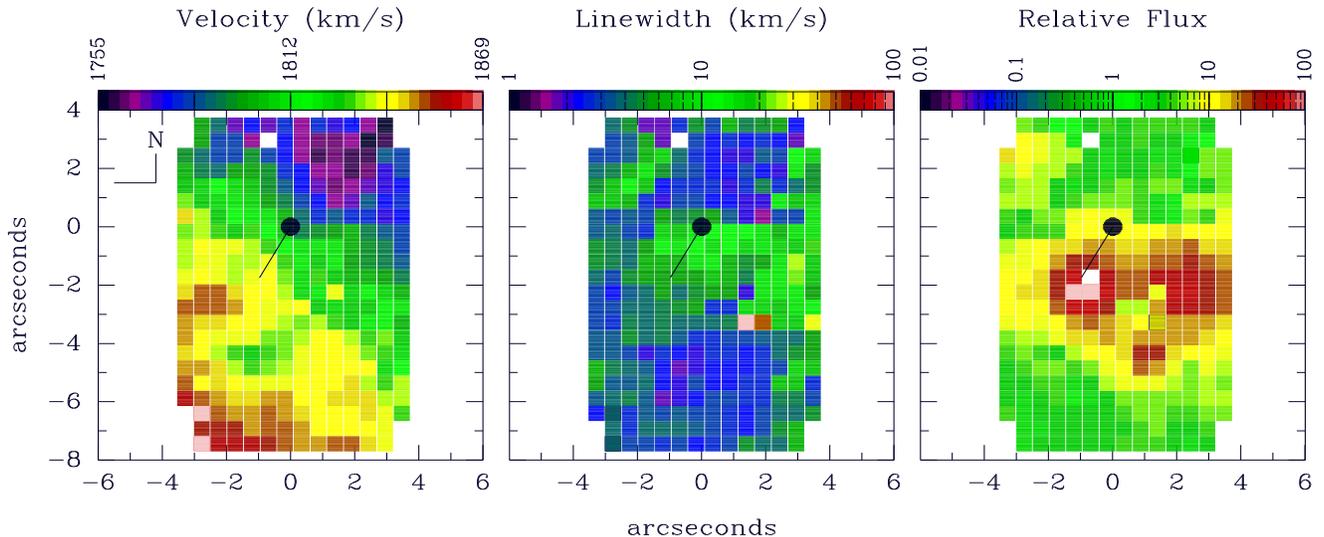}
\caption{Velocities (left), widths (center) and fluxes (right) of the
narrow line profile component (component A). Rotation is clearly
apparent in the velocity field and a simple model fits the velocity measurements
adequately. The location of the best-fit kinematic center is marked with a
filled circle. The line from the point marks the direction of the kinematic
major axis. Approximately 5\arcsec South of the kinematic center is a
significant kinematic asymmetry.
We find nothing remarkable in the width field, perhaps just a slight broadening
of the H$\alpha$ line roughly coincident
with the kinematic center.
Two peaks in the flux distribution are clearly visible, but as we show
in Section \ref{s:reg}, neither peak corresponds to the locations of
the central cluster or the kinematic asymmetry.}
\label{f:comp1}
\end{figure*}

\begin{figure*}
\vbox to 2.8in{\rule{0pt}{6.5in}}
\includegraphics{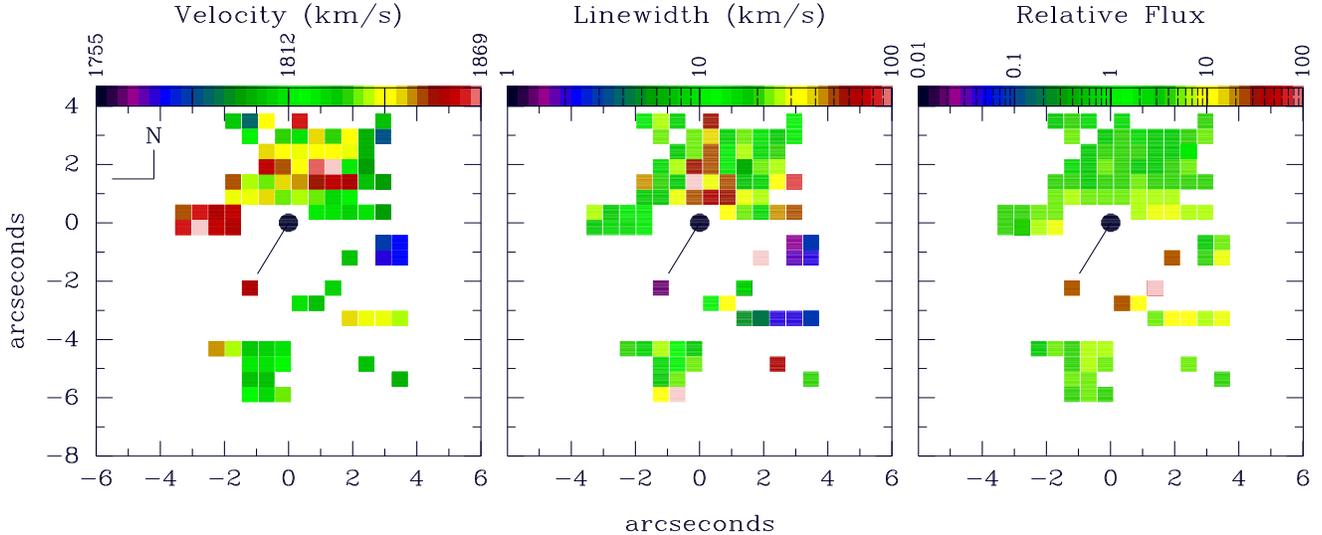}
\caption{Velocities (left), widths (center) and fluxes (relative to Figure 2;
right) of the
second, broader component (component B).
No rotation is apparent and no structure is
apparent in either the width or flux fields. }
\label{f:comp2}
\end{figure*}

Data were overscan- and bias-corrected and trimmed using the NOAO {\sl
IRAF}\footnote{IRAF is distributed by the National Optical Astronomy 
Observatories, which are operated by the Association of Universities 
for Research in Astronomy, Inc., under cooperative agreement with 
the National Science Foundation.}
package {\it ccdproc}. Cosmic ray rejection was performed before
spectral extraction using a method described in Andersen \etal (2006).
Following cosmic-ray cleaning, basic
spectral extraction,
flattening, wavelength calibration and sky subtraction were done using
{\sl IRAF} {\it dohydra}.
Extracted one-dimensional spectra were
field-flattened with dome flats and wavelength-calibrated using ThAr
emission spectra.  Finally, the 14 sky spectra were averaged and the
mean sky spectrum was subtracted from the 299 source spectra.

Once spectra were processed, we identified
H$\alpha$ emission-lines and
measured Gaussian fluxes, widths, centers and the corresponding errors
for lines in a 20\AA\ spectral window around the redshifted H$\alpha$ line.
We detected H$\alpha$ emission in 298 of 299 object spectra.
Most H$\alpha$ emission-lines of NGC 2139 required multiple
Gaussian lines to be fit.  
We modified the algorithm used in Andersen \etal (2006) in order to
incorporate two Gaussian profiles in the Marquardt--Levinson fit.
Errors on line centroids
were obtained from the covariance matrix.
We fit all H$f\alpha$
spectra with both one and two Gaussian profiles.  We compared
the reduced $\chi^2$ ($\chi_\nu^2 = \chi^2/(n-p)$, where $n$ is the number
of data points and $p$ is the number of free parameters)
values of both fits.  Spectra with a $\chi_\nu^2$ 
ratio less than 1.1 were classified as single lines (type 1), 
while all others were classified as having two components. 
Of the 298 detected H$\alpha$ emission lines, 278 were better fit with two 
Gaussian lines. Using a visual inspection,  
we classified the multiple line fits in three categories 
(Figure \ref{f:doubleline}): single narrow 
lines with wide wings (type 2); single narrow lines with wide wings that 
are offset from the central peak (type 3); and finally two kinematically 
distinct features (type 4). 
We find that only 88 of the 278 lines are best fit by two Gaussian profiles
and belong to the last category type 4 with two unique components. 

With this information, the kinematics of NGC 2139 can be divided into 
two components. 
To extract the measures of widths and centroids for component A, we used 
the single Gaussian for type 1 spaxels and the properties of the more highly 
peaked Gaussian for type 2, 3 and 4. For the flux in the line we used the 
flux in the core and the wings, when present, for types 1, 2 and 3, while 
we used the flux in the main component only for type 4 spaxels.  
For the second component, called B, we used the properties of the secondary 
H$\alpha$ emission peak when unambiguously present (type 4). 

The first component (A) shows clear rotation in the velocity field 
(Figure \ref{f:comp1}). No structure in the width of this component 
is apparent, while it shows two distinct peaks in flux. 
On the other hand, the second component (B; Figure \ref{f:comp2}) 
does not rotate, clearly has a larger dispersion in line widths and has no
clearly discernible peak in flux. 
Physically speaking, we have attempted to separate a disc-motion 
component A from a random-motion component B. 
In almost all of the cases the widths of the narrow component A 
is around $\sigma\sim$15 km/s and that of the broad component B is 
$\sigma>$50 km/s. That and  the regularity of component A and 
the irregularity of component B support our interpretation as 
disc and random-motion component, respectively.

\section{Analysis}
\label{ana}

\subsection{Velocity Field Modeling}

We used the centroids of the narrow component A described above
as the input to our velocity field modeling.
Our velocity field model incorporates two simplifying assumptions:
the rotation of the gas in the disk is circular and the shape of the rotation
curve can be approximated by a tanh function [$V_{\rm mod}(R,\theta)=
V_{sys} + V_{rot} {\rm tanh}(R/h_{rot})\cos\theta$; Andersen \etal 2001].
Keeping these assumptions in mind, we are in general able to solve for 
a number of disk parameters, including 1) the kinematic center, 2) 
the major axis position angle, 
3) the systemic velocity, $V_{sys}$, 4) a rotation velocity, $V_{rot}$, and
scale length, $h_{rot}$, which match the velocity field over the ARGUS field.
The best fit and covariance matrix were determined by 
minimizing $\chi^2$ using a modified Levenberg-Marquardt algorithm.

One can compare an IFU spaxel to a radio beam; both smooth or ``smear''
signal velocities within either the spaxel or the radio beam. 
We accounted for the effect of beam smearing by mapping the model over
the region spanned by the observations on a grid with a scale of
10 ``pixels'' per spaxel (0.052\arcsec). 
By determining the standard
deviations of velocities within each spaxel and incorporating this error
in our calculation of $\chi^2$
we are, in effect, applying a beam-smearing correction. 

\begin{figure*}
\vbox to 3in{\rule{0pt}{6.5in}}
\includegraphics{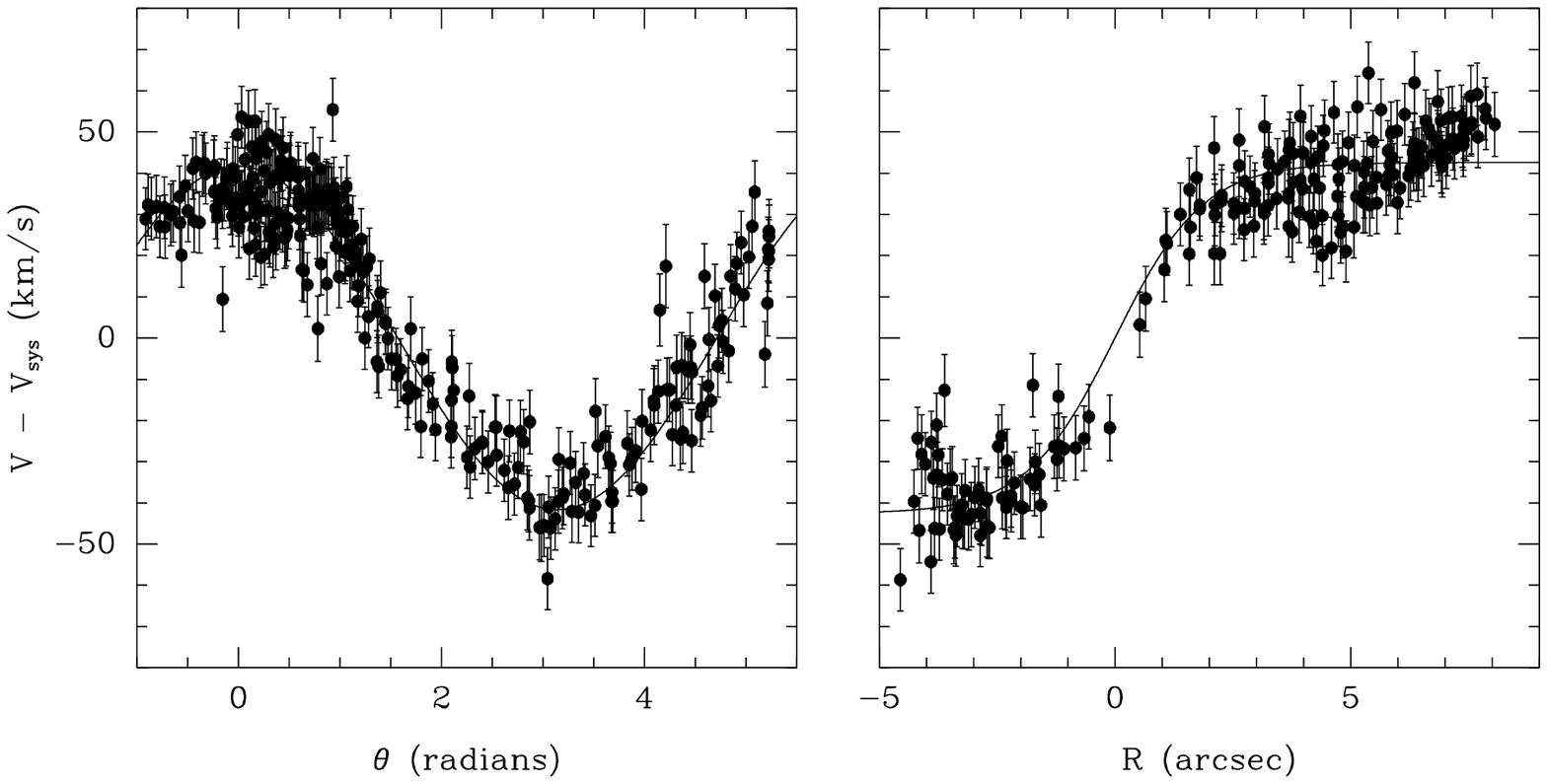}
\caption{Measured velocities versus $\theta$, the de-projected azimuthal angle
(left) and $R$, the de-projected radius (right) for an inclination of 25
degrees. In the rotation curve on the right, only velocities measured within
60$^\circ$ of the major axis are included.  All velocity measurements on this
rotation curve are projected onto the major axis by multiplying by
sec($\theta$).   Error bars on velocities reflect the quadrature sum of the three
terms included in equation 2: centroid measurement errors, $\sigma_{\rm mod}$,
and a beam-smearing contribution.}
\label{f:rc}
\end{figure*}

Our best fits to the velocity field yielded small velocity-field
residuals (Figure 4), with standard deviations of just
$\sim7$ km/s, but our velocity centroiding errors were 
even smaller, resulting in
minimum reduced-$\chi^2$, $\chi^2_\nu$, for these fits which were typically
much greater than
unity ($\chi^2_\nu\approx50$). For any fit where
$\chi^2_\nu$ is much greater than unity there are several possible
explanations: (1) the residuals of our velocity field model
fit are not Gaussian, (2)
error-bars on the line centroids are underestimated, or (3) the model is not
a good representation of the data.
Indeed, one would not expect our simple model to
approximate the true velocity field because in detail the velocity
fields of real galaxies can be complicated on fine spatial scales.
Since the spectral and spatial resolution of our ARGUS observations is 
relatively high, it is reasonable
to suppose we are seeing the true complexities of the kinematics
at the center of NGC 2139.
Presumably a class of models exists
which would better match the data, but would become increasingly complex.

Since we are not interested here in modeling the random
motions of HII regions and the non-random, high spatial frequency
streaming associated with spiral arms or other similar features,
we assumed these variations have a random spatial distribution.
We compensated for these contributions by adding
a ``fuzziness'' term to our model. Specifically, we follow
the probability theory arguments of Rix \etal (1997) and add
an extra error term, $\sigma_{\rm mod}$, into the $\chi^2$ sum
(which we denote as $\chi^2_\delta$ to differentiate it from the
usual definition of  $\chi^2$):
\begin{equation}
\chi^2_\delta = \sum_i {{(V_{{\rm mod}_i} - V_{\rm{obs}_i})^2}\over{
\sigma_{\rm mod}^2+\sigma_{{\rm obs}_i}^2}},
\label{eq:rix}
\end{equation}
where $V_{{\rm mod}_i}$ is the model velocity at the location of the
$i^{th}$ spatial element,
$V_{{\rm obs}_i}$ is the observed velocity of the $i^{th}$ spatial element,
and $\sigma_{{\rm obs}_i}$ is the standard deviation on $V_{\rm{obs}_i}$,
including both the effects of beam-smearing and the centroid measurement
error.

Using our model, we were not able to find a unique solution 
for the disk inclination of NGC
2139, but we were able to find unique solutions by fixing the inclination in
the fit.  The inclination derived from inverting the Tully-Fischer relation
(Rix \& Zaritsky 1995)
falls between 25$^\circ$ and 40$^\circ$. Therefore, 
we fit the velocity field with fixed
inclinations varying between 5 and 50 degrees, normalizing the 
reduced-$\chi^2_\delta$ to unity (within 1\%) by choosing an appropriate 
value of $\sigma_{\rm mod}$.
We found a major axis position angle of 150$^\circ\pm1^\circ$,
an observed rotation velocity of $V_{rot}=44\pm2$ km/s, a heliocentric 
systemic velocity $V_{sys}=1793\pm1$ km/s, and a scale length 
$h_{rot}=1.8^{\prime\prime}\pm0.2^{\prime\prime}$ ($\sim 200\pm20$ pc).
The error on the kinematic center was just $0.15^{\prime\prime}$ ($\sim 20$ pc).
While $\sigma_{\rm mod}$ systematically varied as a function of inclination, 
the other kinematic parameters and their associated errors, especially the 
kinematic center, do not change significantly with respect to the fitting 
error. 

The isovelocity contours are not bent strongly as in classic disk
galaxy ``spider diagrams,'' as evidenced by Figure \ref{f:comp1}.
While this adds to the uncertainty on the exact location of
the kinematic center along the North-East to South-West axis of the galaxy,
the spacing between contours (tightly bunched near the best-fit center and
increased space between isovelocity contours further out) implies that the 
kinematic center robustly lies close to the nominal minor axis 
(compare Figure \ref{f:rc}).

The kinematic center is offset with respect to the peaks in H$\alpha$ flux
and, as we shall see in Section \ref{s:reg}, from the central cluster as well. 
The high precision with which we determine the kinematic center
from our velocity field fitting does not include possible sources of
systematic uncertainty:
(a) We extracted one H$\alpha$ component from two that were at different
velocities without being able to unambiguously identify a physical
reason to do so. (b) Considering only this one component used
in our velocity field modeling, there is a kinematic asymmetry 
5$^{\prime\prime}$ along the major axis which is not understood 
(Figure \ref{f:comp1} and Section \ref{s:reg}). 
If this was taken as an indication that 
the velocity field is not well ordered, it could be used to argue 
against the NC formation mechanism that requires the center to be a 
special place in these late-type galaxies. 

\begin{figure*}
\vbox to 2.35in{\rule{0pt}{3.5in}}
\includegraphics{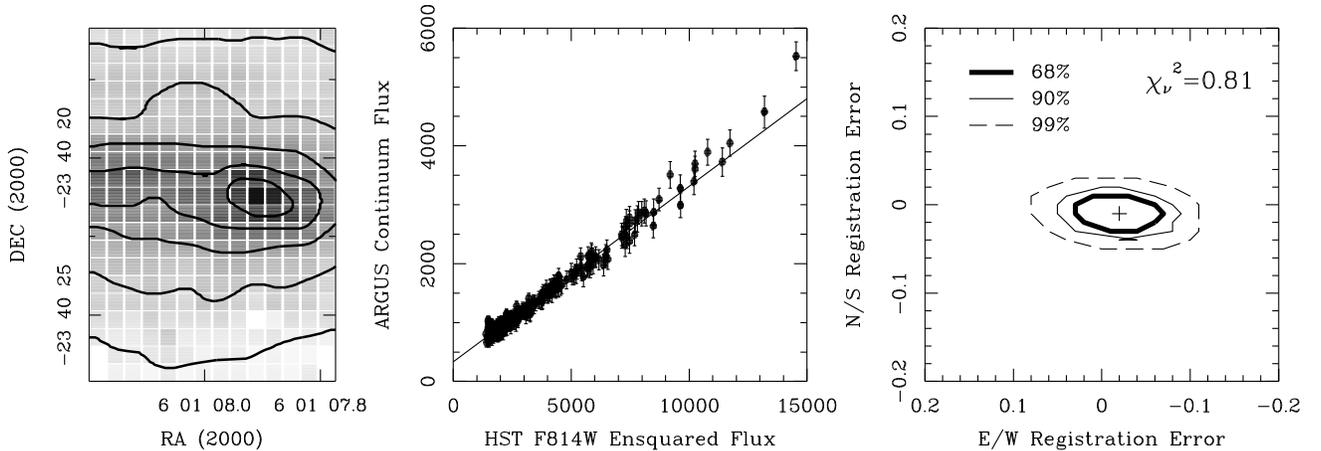}
\caption{ Left Panel: continuum levels from VLT ARGUS spaxels overlayed with
smoothed (by 0.8\arcsec) HST image contours of the
center of NGC 2139 at the best-fit registered position.
Center Panel: spaxel-ensquared HST WFPC2 F814W flux versus
ARGUS continuum flux with best fit linear regression.
Right Panel: error contours of the image to IFU registration based on
the $\chi^2$ map.  The reduced $\chi^2$, $\chi^2_\nu$, for the fit was
0.81, indicating a good fit.}
\label{f:register}
\end{figure*}

On the other hand, the center we find is quite robust. Leaving out some 
of the more questionable data, removing the kinematically asymmetric data, 
and trying to force the code to produce a center at a different location
by varying the initial fit parameters
all fail to produce a significant shift in the model kinematic center;
none of the kinematic parameters change more than the errors quoted above.
So, assuming a simple rotation curve and velocity field model,
we find that that the best fit kinematic center is well-constrained
which implies a single strong minimum in $\chi^2$ space.
Figure \ref{f:rc} shows the derived rotation velocities against 
de-projected azimuthal angle and radius. We recover the expected 
sinusoidal shape and s-shape 
respectively, indicating that our simple model for the velocity field does
an adequate job of describing the data.  Forcing the kinematic center away from 
the best fit location would significantly alter the shape of these plots.

Further supporting evidence for the accuracy and
interpretation of the ARGUS velocity field comes from the HI recession velocity:
at first glance, the kinematic parameters derived from the H$\alpha$ velocity
field do not agree with the HI observations.  In particular the 
H$\alpha$-derived systemic 
velocity (1793$\pm1$ km/s) is much smaller than the value
1837 km/s found from HIPASS.  While global
line profile asymmetries can create differences between HI single dish 
recession velocities and the systemic velocities derived from velocity fields
(Andersen \& Bershady 2008), the sense of the discrepancy is reversed.  The
true systemic velocity, based on the asymmetry of the HI profile, should
be even higher than 1837 km/s. However, as mentioned in Section \ref{s:prev}, 
the Condon \etal (1996) radio continuum maps
suggest that there are two discrete sources of radio emission 
from NGC 2139. If the stronger peak in the HI profile corresponds to the
larger source of radio continuum emission associated with the main body
of NGC 2139, then we find good agreement between this peak velocity 
($\sim1795$ km/s) and our result.  Furthermore, if the HI emission is coming
from two distinct sources at two distinct velocities, then our measured
rotation velocity will account for almost the whole width of the primary
peak.  Even if this ``two gas clouds'' picture is not correct, we are observing
a sizable amount of the total rotation in NGC 2139 (44 km/s from
our H$\alpha$ velocity field versus 94 km/s from the double-peaked 
HI profile) in a
very small area near the center of the galaxy, lending credence to our 
interpretation that the kinematic center lies in the ARGUS field of view.

Finally, the velocity field possibly indicates that the luminous 
band described in Section 2 may be a star formation filament 
rather than a stellar bar because, as we show in the next 
section, the star formation knots are indeed located in this band, 
but the velocity field is not distorted by this feature.  The
kinematic center of NGC 2139 is well-separated from this band, and 
while bars are observed offset from galaxy centers in late-type 
galaxies, such as the LMC (de Vaucouleurs \& Freeman 1972),
this behavior is often attributed to tidal forces (e.g., between the Milky 
Way and the LMC; van der Marel \etal 2002).  
Observations provide some evidence that NGC 2139 is affected 
by tidal forces, but whether the possible tidal effects are 
quantitatively sufficient to explain the offset we see between the
stellar and gas kinematics is beyond the scope of this paper.
Although we note that the distortions we observe in the velocity 
field seem unlike those expected from a bar 
(Roberts, Huntley \& van Albada 1979), a separate study would 
be needed to convincingly argue against the presence of such 
a bar. 

\subsection{IFU continuum to HST Image Registration}
\label{s:reg}

Before we can compare our best fit kinematic center to the location of the
central cluster in NGC 2139, we need to 
register the HST WFPC2 F814W image of Boeker \etal (2002)
to the continuum flux from the 
ARGUS IFU. We first measured continuum levels 
from the IFU data within a 400 \AA\ spectral window between 6600 {\AA} $<
\lambda < $ 7000 \AA\ in which emission lines were masked. 

We registered this continuum IFU image to the HST image by first smoothing
the HST image with a 0.8\arcsec\ Gaussian to roughly match the seeing
during our ARGUS observations.  Then, for a given position of the IFU 
with respect to the smoothed image, we extracted the flux within the 
footprint of each IFU spaxel.  We fit a linear relation between the
IFU and the HST spaxel fluxes and tabulated $\chi^2$.  We mapped $\chi^2$
over a grid in North-South and East-West offsets of the IFU with respect to the
HST image, and found a very good fit between the HST extracted fluxes and IFU 
continuum fluxes. We used our $\chi^2$ map to generate errors on our
image to IFU registration (Figure \ref{f:register}).

\begin{figure*}
\vbox to 5.75in{\rule{0pt}{6.5in}}
\includegraphics{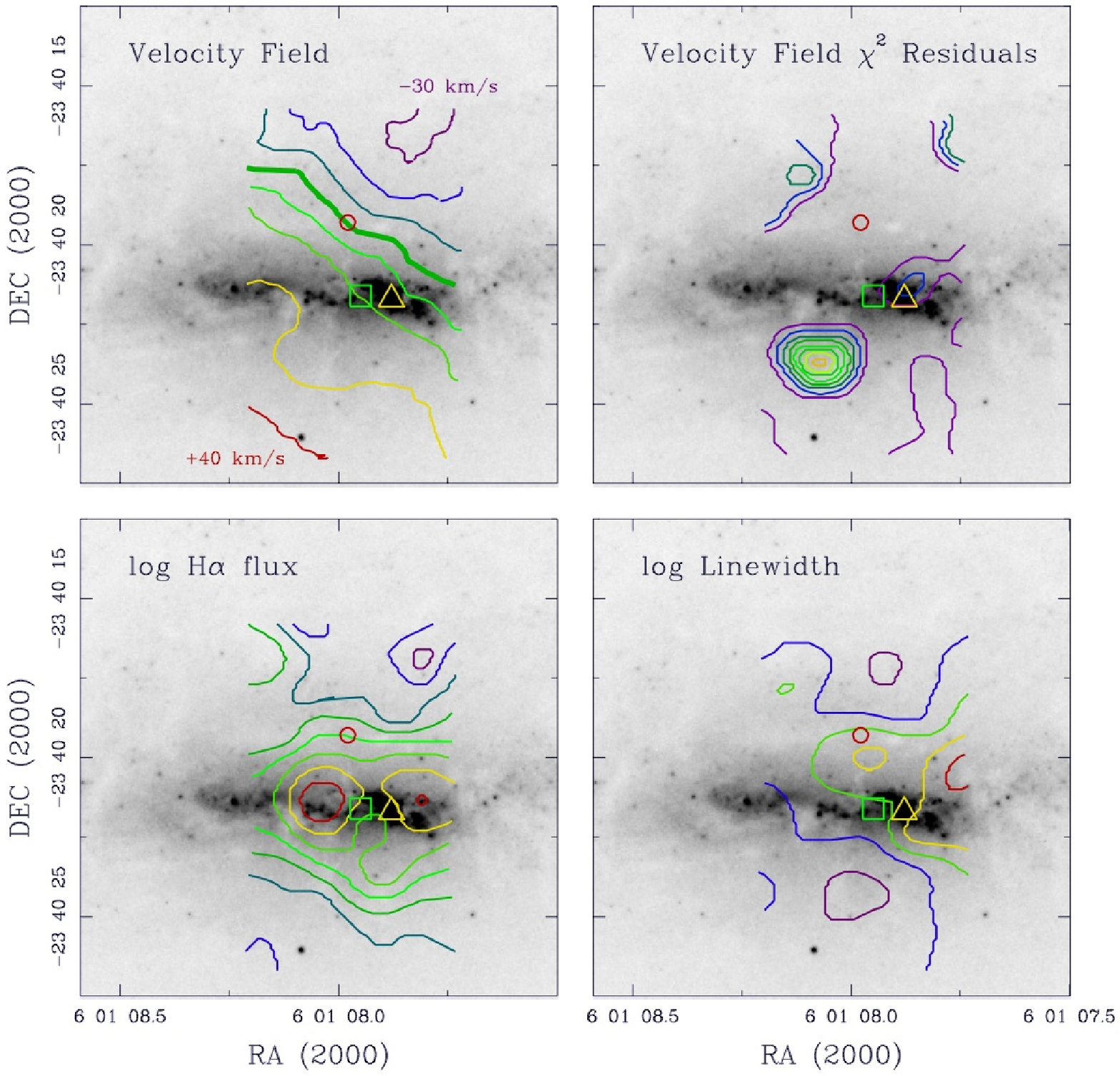}
\caption{Upper Left Panel:  H$\alpha$
isovelocity contours for the central regions of NGC 2139 overlayed
on the HST WFPC2 F814W image.  In all plots of this figure,
the red circle marks the location of the kinematic
center, the green box marks the location of the photometric center,
and the triangle marks the location of the central cluster.
All coordinates refer to those of the HST image. Isovelocity contours
are separated by 10 km/s, with the extreme velocities listed with respect
to the systemic velocity.
Upper Right Panel: $\chi^2$ contribution to the velocity field best-fit kinematic
model. Different
colors are separated by $\Delta \chi^2=1$, from $\chi^2=1$ (purple) to
8 (red).   The large kinematic asymmetry 5\arcsec\ from the kinematic center is
the dominant feature and does not appear to be associated with any feature in
the HST image. A second significant kinematic asymmetry is apparent, coincident
with the central cluster.
Lower Left Panel: H$\alpha$ flux level contours (see Figure \ref{f:comp1}
for relative values of the fluxes) overlayed on the HST image.  Two strong
peaks on star formation are apparent.  Neither is coincident with the
central star cluster.  In general,
the luminous band has higher H$\alpha$ fluxes than the regions outside.
Lower Right Panel: H$\alpha$ linewidth contours overlayed on the HST image.
Each contour is separated by steps of 0.15 in log linewidth from
$10^{1.05}=11.2$ km/s (red) to $10^{0.45}=2.8$ km/s (violet).
While the linewidths are slightly larger near the star cluster and within
the luminous band, we do not see a clear correspondence between the
linewidths and any of the other mapped quantities.
}
\label{f:vfield}
\end{figure*}

With the IFU registered to the HST image, we can assign 
coordinates to the best fit kinematic center.  This kinematic center is
located at RA = 06$^h$01$^m$07.98$^s$ and DEC = 
-23\arcdeg 40\arcmin 19.3\arcsec\ (J2000) with
errors of $\sim$0.2\arcsec\ on each axis (based on both registration errors and 
the formal velocity field errors).  This is 2.8\arcsec\ away from the 
location of the central cluster 
at RA = 06$^h$01$^m$07.88$^s$ and DEC = -23\arcdeg 40\arcmin 21.7\arcsec\
(Figure \ref{f:vfield}).  
Here and
hereafter all (RA,DEC) values refer to the J2000 coordinate system of
the WFPC2 F814W HST image, which has an absolute accuracy with respect to
the ICRS system of $\sim 1$\arcsec.

\begin{figure*}
\vbox to 4.5in{\rule{0pt}{4.5in}}
\includegraphics{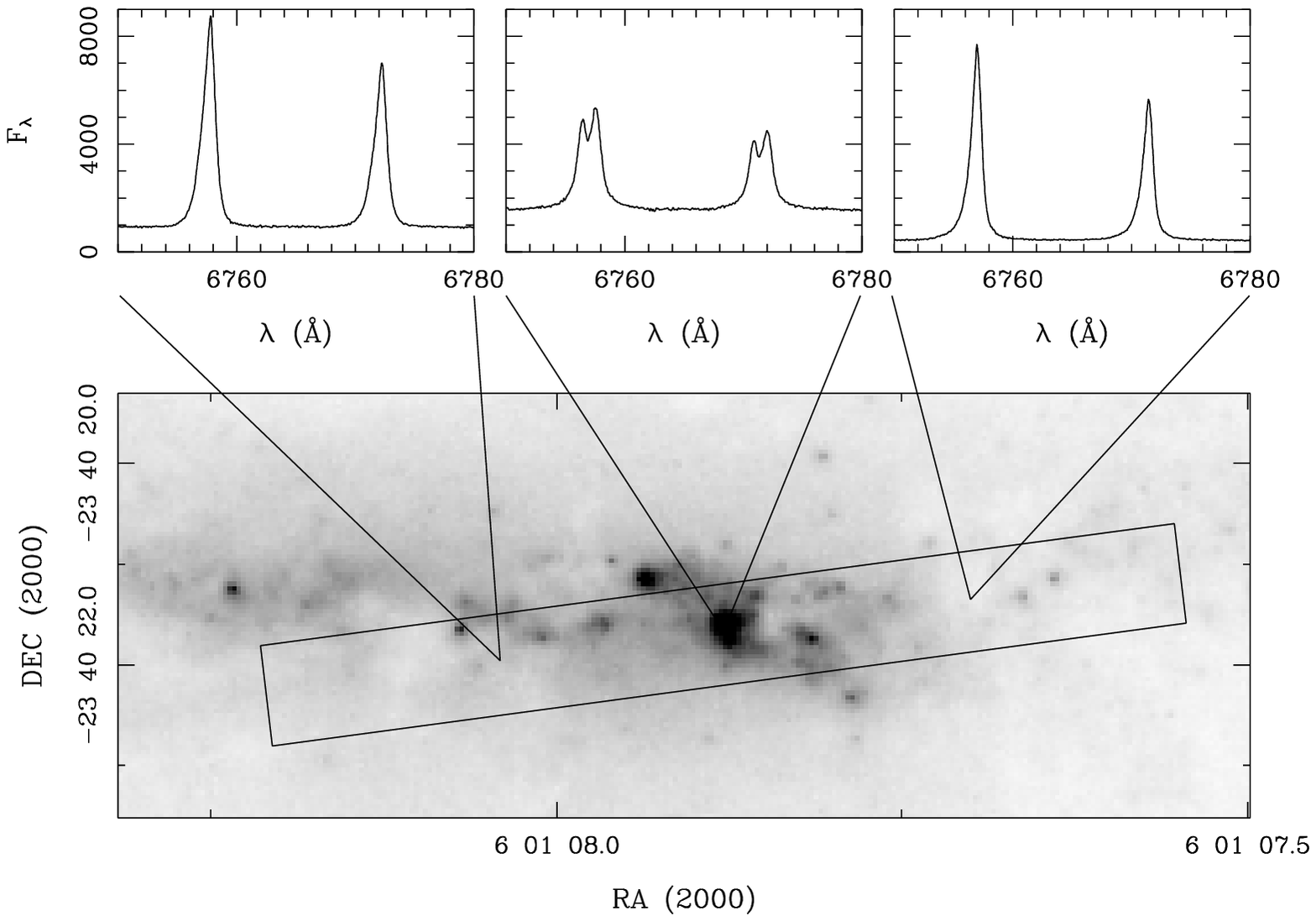}
\caption{Top Panels: the spatial profiles of the
[S II] emission lines measured
from VLT/UVES high resolution spectra (R $\sim$ 30000) in three apertures
of 1\arcsec $\times$ 1\arcsec. One aperture includes the central cluster itself
and one aperture 2 arcseconds on either side are also shown.
Bottom Panel: HST WFPC2 F814W image centered on the central star cluster
of NGC 2139 with the UVES slit overlayed.
The broad emission lines in general and
the double peaked emission line on the central cluster
in particular are compatible with
a very young massive star cluster (SSC) in a region actively forming stars.}
\label{f:apertures}
\end{figure*}

We note that the central cluster 
is coincident with a region poorly fit by the global 
velocity field (Figure \ref{f:vfield}), although the largest
deviation from the velocity field does not correspond to any feature in
the image. We also find that the central cluster is located near,
but is not coincident with, the two peaks in H$\alpha$ flux. The
central cluster 
is separated by 3.2\arcsec\ from the brightest source of H$\alpha$ emission.
This brighter, primary peak of H$\alpha$ emission has 
no clearly identifiable counterpart in the HST I-band image. 
We postulate that the primary peak in H$\alpha$ emission represents an 
even younger SSC perhaps still enshrouded in its birth-cloud. 
Perhaps future mid-infrared observations could reveal structures 
coincident with the large deviation from the velocity field
South-West of the central cluster and the bright peak in H$\alpha$ emission
to its West.

\subsection{Determination of the Photometric Center}
\label{s:photocent}

Finally, we re-determine the location of the photometric center of 
NGC 2139 from the HST WFPC2 F814W image. This exercise had been carried 
out in B\"oker \etal (2002) with the aim to derive photometric 
centers for a large number of galaxies in a homogeneous way. Here we 
focus on understanding one special case. Indeed, NGC~2139 has a 
rather irregular appearance on the PC chip. In particular, 
the luminous band going through the center of the object has a number 
of secondary brightness peaks (secondary to the central cluster), 
which have the effect 
to shift the photocenter each time the radius of the isophotal ellipse 
becomes large enough to include one more of these peaks. We have therefore 
convolved the image with a $\sigma = 10$\arcsec\ Gaussian before using the 
exact same setup as B\"oker \etal (2002). We now define the isophotal 
center to be the arithmetic mean of all ellipse centers between radii of 
0.7\arcsec\ and 18.4\arcsec. We thus obtain the following photocenter: 
RA = 06$^h$01$^m$07.95$^s\pm$0.09$^s$ and DEC = -23\arcdeg 40\arcmin 
21.6\arcsec$\pm$0.4\arcsec\footnote{The photometric center determined 
from the DSS image of NGC 2139 was statistically equivalent.}.
The photometric center is 2.3\arcsec$\pm$1.4\arcsec\
away from the kinematic center and 
1.0\arcsec$\pm$1.4\arcsec\ away from the central cluster.
We note that this result still depends strongly on the range 
of radii used to determine the photometric center. 
The location of the photometric center is shown in Figure \ref{f:vfield}.

\section{Nuclear Cluster or young Super Star Cluster?}
\label{s:evidence}

We have provided new data to judge whether the classification of the central star
cluster in NGC 2139 as a NC is correct.  While all the results we
have obtained from the H$\alpha$ velocity field are consistent with
the interpretation of the cluster being a nascent NC, we examine
here the possibility that this cluster is not a NC and instead is a
more normal young super star cluster (SSC).
This is supported by the emission line profiles as observed in the
VLT/UVES data published in Walcher et al. (2005, 2006). 
Figure \ref{f:apertures} shows the [S II] doublet emission lines in 3
apertures along the spatial direction. It is clear that the
lines are broadest at the location of the central cluster. The FWHM for 
this aperture, as determined from a single Gaussian fit is 2.3 {\AA} = 100 km/s.
However, the lines have two separate kinematic components, consistent with
the ARGUS H$\alpha$ data.
These two components can be fit by two Gaussians, but 
high-velocity wings remain that are not fit even by this
2-Gaussian fit. These wings have a full width at zero intensity of 300 km/s.
The redder peak is coincident in velocity with the cluster stars (measured
from the Ca Triplet), while the bluer peak is separated by
approximately 55 km/s. One could speculate that we are seeing a
super-bubble around the cluster, where the part that is coming towards
us is visible, while the receding part of the bubble is hidden by
extinction inside the cluster itself. 
Such complex emission lines with two components are routinely seen in 
HII regions and very young clusters (e.g. Vanzi et al. 2006, Henry et al. 2007).

The stellar velocity dispersion for the cluster in NGC 2139 is
$\sigma_* = 16.5 \pm$ 1 km/s (Walcher \etal 2005).
So its escape velocity is approximately $\sqrt{2}\sigma_* \approx 23$ km/s.
This means that the gas in the peak coincident with the cluster
is marginally bound ($\sigma_{\rm gas} = 20 $ km/s). However, neither the
high-velocity wings, nor the gas moving towards us are bound to the
central cluster.

These properties can be compared to the discussion in Gilbert \& Graham
(2007). These authors publish emission line widths for a large sample
of SSCs in the Antennae. The velocities in the NGC 2139 cluster are 
consistent with, yet at the high end, of those
observed by Gilbert \& Graham. As these authors show, such high
velocities and complex line profiles are associated with winds
and mass loss from very young star clusters, so-called SSCs.
Naturally, star formation and winds could also be 
associated with a NC undergoing a rejuvenation burst.

We also remind the reader that the effective radius of the NGC 2139 cluster 
is 10pc and is one of the largest in the B\"oker et al. (2004) sample.
This could be interpreted in the context of the study of Mengel \etal (2005), 
who show that the radii of SSCs are linked to their age. These authors show 
that the effective radii of typical SSCs are 16$\pm$15pc for 4 Myr old clusters 
and 6.5$\pm$5.3 pc for 8-11 Myr old clusters. 
In Gilbert \& Graham these large radii are interpreted
as expansion following mass-loss through winds. Of course, a similar
effect could puff up the radius of young NCs when compared to their older
cousins.

The environment around the central star cluster also suggests that this may
be one of multiple SSCs in NGC 2139.  The multiple sources of radio continuum
emission, the luminous band in which the central star cluster
is located, 
the larger-scale tidal features and 
the multi-component H$\alpha$ and [S II] emission
all are consistent with a recent merger event, which may drive gas to the center
and initiate the formation of SSCs.  Indeed, the two bright H$\alpha$
sources in the vicinity of the central star cluster may be very young SSCs.

While the above arguments lead to a certain degree of uncertainty with
regards to the classification of the central cluster in NGC 2139, we can
predict whether we would identify the cluster 3 Gyr from now as a NC.
Using a mass-to-light ratio
M/L$_I$=0.87, as predicted by and model predictions from 
Bruzual \& Charlot (2003) for a 3 Gyr old
stellar population and using a cluster mass of $8\times10^5$ M$_{\odot}$, we
obtain L$\approx 7\times10^5$ L$_{\odot}$ after 3 Gyr.  This is
significantly brighter than the cutoff luminosity of a NC as observed 
in B\"oker \etal (2002)\footnote{No NC with a luminosity below M$_I$=-9 
was detected, despite sufficient sensitivity. Using
M$_I$ = -2.5 log(L$_I$) + 4.08 this corresponds to $2\times10^5$ L$_{\odot}$.}.

If the identified central star cluster can somehow merge
with other nearby clusters (such as those associated with the two bright
peaks of star formation in the H$\alpha$ map) and fall to the kinematic
center, then
the resultant NC could be even brighter. If this were to happen, this would
support the NC formation scenario whereby multiple 
stellar clusters coalesce to form the final NC (Fellhauer \etal 2002). 
That so many young clusters are present near the center of NGC 2139 is
another indication that the galaxy recently experienced a tidal interaction
capable of forming clusters (Bekki \etal 2004).

We now assess the probability that such a massive and luminous 
cluster would form by chance in a late-type disc galaxy such as NGC 2139. 
To that end, we compute the star formation rate (SFR) necessary to produce 
a cluster of mass similar to the central cluster in NGC 2139. Due to size of 
sample effect, the maximum mass $M_{max}$ of a star cluster scales with the 
total number $N$ of star clusters formed, or equivalently with the star 
formation rate (see Larsen, 2002). The central cluster in NGC 2139 has an absolute 
I-band magnitude of M$_I$ = -12.65 (B\"oker \etal 2002) and from its age 
and Bruzual \& Charlot (2003) we infer a colour V-I = 0.5. Following the 
relations given in Weidner \etal (2004), we infer that a total SFR 
of roughly 1 M$_{\odot}$/yr is needed to form 
one such cluster. Emission line fluxes for NGC 2139 are 
given in Moustakas \& Kennicutt (2006) for a nuclear aperture of 2.5\arcsec $\times$
2.5\arcsec\ and the total galaxy (120\arcsec $\times$ 120\arcsec). 
The H$\alpha$ fluxes are $23.11\pm0.97 \times 10^{-15}$ ergs s$^{-1}$ cm$^{-2}$ and 
$3310\pm130 \times 10^{-15}$ ergs s$^{-1}$ cm$^{-2}$ respectively. Using a distance 
of 23.6 Mpc (B\"oker et al. 2002) and the standard Kennicutt (1998) conversion 
of SFR [M$_{\odot}$/yr] = 7.9 $\times$ 10$^{-42}$ L$_{H\alpha}$ [erg/s],
we obtain nuclear and total SFRs of 0.01 and 1.7 M$_{\odot}$/yr, respectively. 
The central cluster 
is the brightest cluster in the area of the PC chip, which is 36\arcsec 
$\times$ 36\arcsec\ (4kpc $\times$ 4kpc). 
However, the surface brightness of the galaxy falls rapidly 
outside of the chip, so it is likely that most of the H$\alpha$ flux observed 
by Moustakas \& Kennicutt stems from the actually observed area. 
In summary, it is likely that a cluster similar to the one observed in NGC 2139 
forms somewhere in the disk of the galaxy. The likelihood that this happened by 
chance in the central 2.5\arcsec $\times$ 2.5\arcsec\ however remains low. 
Note though, that with the 40 km/s current rotational speed of the galaxy
and the estimated cluster age, the cluster could have moved by as much as 1.5 
kpc since formation. As the cluster seems to have formed from disk material 
and is still associated with the gas over-density in which it formed, it is 
not likely, however, that the cluster has a large peculiar motion with respect 
to the rest of the galaxy disk.
Whether special conditions close to the center of the galaxy need to be 
invoked thus remains open.

\section{Summary}
\label{s:summary}

We have presented VLT/ARGUS IFU data to ascertain
whether the location of the kinematic center of the late-type 
galaxy NGC 2139 and the location of its central cluster, previously 
classified as a NC, coincide spatially.
We have analyzed the velocity field in the ARGUS field of view 
and found the kinematic center to be at  06$^h$01$^m$07.98$^s$ and DEC = 
-23\arcdeg 40\arcmin 19.3\arcsec\ with
errors of $\sim$0.2\arcsec\ on each axis.
The location of the central cluster is at RA = 06$^h$01$^m$07.88$^s$ and DEC = -23\arcdeg 
40\arcmin 21.7\arcsec, 2.8\arcsec\ offset from the 
kinematic center. The photometric center is at RA = 06$^h$01$^m$07.95$^s$ 
$\pm$0\fs09 and DEC = -23\arcdeg 40\arcmin 
21.6\arcsec$\pm$0.4\arcsec\ and is 2.3\arcsec$\pm$1.4\arcsec\
away from the kinematic center and 
1.0\arcsec$\pm$1.4\arcsec\ away from the central cluster.
While some caveats remain concerning the precision which we quote for the location 
of the kinematic center, it appears unlikely that the 
location of the NGC 2139 central star cluster is coincident with the kinematic
center.

Based on the peculiar H$\alpha$ and [S II] emission from the core of NGC 2139,
we conclude that the central cluster is actively forming stars. While the 
properties of the central cluster in NGC 2139 are thus compatible with 
those of a SSC, we have presented several arguments why it can 
nonetheless be considered to be a NC progenitor object: 
\begin{itemize}
\item Based on the projected luminosity of the cluster in 3 Gyr, it would 
be classified as a NC in the survey of B\"oker et al. (2002), if the cluster 
does not lose a significant fraction of its mass due to evaporation. 
\item Other young star clusters associated with peaks in the H$\alpha$
distribution may
merge with the visible one to produce a multi-aged, 
very luminous NC. Alternatively, more gas from the luminous band
may fall into the central star
cluster and form stars in situ.
\item While the NGC 2139 central cluster is not presently
at the center, the relatively
well-ordered velocity field means that the cluster should fall to the center
within a relatively short period of time:
if the cluster is 320pc (2.8\arcsec) 
away from the kinematic center with a rotation velocity of 15 km/s, Equation
\ref{eq:milos} yields a derived dynamical friction timescale of 110 Myr,
comfortably longer than the age of 41 Myr derived by Walcher \etal (2006). 
It should be noted, however, that this is only an order of magnitude estimation 
as it will depend on the exact form of the gravitational potential. In the 
case considered by Milosavljevi\'c (2004), the value we derive is a lower 
bound. On the other hand, 
if NGC 2139 is undergoing a dissipative tidal interaction, as the
images and radio data seem to indicate, then the dynamical friction 
timescale for the central star cluster could be significantly shorter 
(e.g. Pe\~narrubia, Just \& Kroupa 2004; Capuzzo-Docetta \& Vicari 2005; 
Miocchi \etal 2006; Fujii \etal 2007).
Regardless, it appears that within a relatively short time this cluster may 
take up residence at the nucleus of NGC 2139.
\end{itemize}

Although we caution that NGC~2139 is in many ways an unusual 
object among all late-type spirals containing NCs, we can 
attempt to view our results 
in terms of the three hypotheses described in Section \ref{s:intro}.
We find a confusing picture: 
the properties of the central cluster in NGC 2139 are compatible with 
the free-roaming seed cluster formation hypothesis (2). 
However, we observe clear rotation of the galaxy, 
which is as predicted by scenario (1).
Finally, a second cluster seems to be forming close to the visible 
central cluster suggesting that for NCs the merger mechanism described 
in scenario (3) can also play a role. 

In summary, it appears that seed NCs have been formed in NGC 2139, 
fed by an inflow of gas, possibly due to a recent merger event, 
It is clear that larger datasets will be needed to 
clarify whether NCs in general are true galaxy nuclei. IFUs with 
larger FOVs would help in this endeavor, particularly if the central
velocity fields are more disordered than that observed in NGC 2139.
The field would also enormously benefit from targeted simulations 
assessing the viability of the three formation scenarios in relation
to all the observed properties of NCs.

\section*{Acknowledgments}       

CJW is supported by the MAGPOP Marie Curie EU Research and Training Network.
JS thanks Andrea Gilbert for informative conversations regarding SSCs.

\bibliographystyle{apj}

\end{document}